\documentclass{iopconfser}
\usepackage{amstext,amsfonts,amsmath}
\usepackage{xcolor}
\newcommand{\be}{\begin{equation}}
\newcommand{\ee}{\end{equation}}

\def\bea{\begin{eqnarray}}
\def\eea{\end{eqnarray}}

\begin{document}

{\rightline{IFT-UAM/CSIC-24-137}}

\title{Noether-Wald and Komar charges in supergravity, fermions, and Killing
  supervectors in superspace}

\author{ Igor Bandos$^{1,2}$, Patrick Meessen$^{3,4}$ and Tom\'as Ort\'{\i}n$^{5}$}

\affil{$^1$ Department of Physics and EHU Quantum Center\\
~~University of the Basque Country UPV/EHU, P.O. Box 644, 48080 Bilbao, Spain}
\affil{$^2$ IKERBASQUE, Basque Foundation for Science, 48011, Bilbao, Spain}
\affil{$^3$ Departamento de F\'{\i}sica, Universidad de Oviedo\\
~~C/ Leopoldo Calvo Sotelo 18, E-33007 Oviedo, Spain}
\affil{$^4$ ICTEA, C/ de la Independencia 13, E-33004 Oviedo, Spain}
\affil{$^5$ Instituto de F\'{\i}sica Te\'orica UAM/CSIC\\
~~C/ Nicol\'as Cabrera, 13--15,  C.U.~Cantoblanco, E-28049 Madrid, Spain}

\email{Igor.Bandos@ehu.eus}

\begin{abstract}
The supersymmetry properties of Killing vectors  and spinors in supergravity theory can be clarified by relating them to Killing supervectors in the  supergravity superspace.  In the superspace approach it is manifest that supersymmetry 'mixes' a Killing vector with its fermionic spinor 'superpartner' and the Killing equations with the generalization of the Killing spinor equations. The latter reduces to the standard Killing spinor equation, albeit with a fermionic spinor, when the fermionic fields are set to zero.  Using these supersymmetry  transformations in the spacetime component approach, we construct a Noether-Wald charge of ${\cal N}=1$, $D=4$ supergravity with fermionic contributions which is diff-, Lorentz- and supersymmetry-invariant (up to a total derivative). The Killing supervector formalism for the maximal $D=11$ supergravity and some related issues are also discussed.
\end{abstract}

\section{Introduction}

The concept of symmetry remains to be one of the most important concepts in modern theoretical physics.
Recently much attention was attracted by the discovery that the very definition of symmetry in quantum theory allows for generalizations (see \cite{Bhardwaj:2023kri} for a very recent review and
an extensive list of references).  The set of generalized symmetries includes higher-form symmetries
\cite{Gaiotto:2014kfa,Gomes:2023ahz}, higher group structures \cite{Sharpe:2015mja,Benini:2018reh}, non-invertible symmetries \cite{GarciaEtxebarria:2022vzq,Schafer-Nameki:2023jdn} and subsystem symmetries \cite{Seiberg:2020wsg}.

The study of generalized symmetries stimulated the interest in unusual conserved charges also in General
Relativity (see e.g.  \cite{Gomez-Fayren:2023qly} and refs. therein) and more
general gravity theories in 4 and higher dimensional spacetimes (see
e.g. \cite{Benedetti:2021lxj,Benedetti:2023ipt,Hull:2024mfb} and
refs. therein), as well as a quest for a deeper comprehension of more standard
types of charges and symmetries, particularly in theories of gravity.

In this direction is oriented, in particular, our recent paper \cite{Bandos:2023zbs} which is devoted to the supersymmetric properties and the fermionic contributions to the Noether-Wald and Komar charges in supergravity. In \cite{Bandos:2023zbs} we show that a generic Killing vector describing a symmetry of a supersymmetric solution of $\mathcal{N}=1, D=4$ supergravity in the presence of non-vanishing fermionic
fields, forms a supermultiplet with a fermionic spinor (called a generalized Killing spinor) both of which are necessary to describe the supersymmetry leaving invariant said solution. The generalized Killing spinor contributes to the supergravity Killing vector equation and obeys a generalization of the Killing spinor equation, which in its turn forms a supermultiplet with the Killing equation.
These results were deduced from the concept of a Killing supervector \cite{Buchbinder:1995uq} and then used to construct the diff-, Lorentz- and supersymmetry-invariant Noether-Wald-Komar charge of $N=1$, $D=4$ supergravity with active spinors, thus completing the construction of  said charge in \cite{Aneesh:2020fcr}.

Our approach provides a solid basis for constructing the supersymmetric thermodynamics of black holes with fermionic (gravitino) hair.  Such a  thermodynamics will be especially relevant in the case of more complicated supergravity theories, ${\cal N}\geq 2$ (AdS) supergravity where black-hole solutions with gravitino hair do exist \cite{Aichelburg:1981rd,Gueven:1982tk,Aichelburg:1983ux}, as well as in higher-dimensional supergravity theories.  In this respect our results in \cite{Bandos:2023zbs} play the role of a proof of concept, while their generalization to the case of ${\cal N}=2$, $D=4$
supergravity will be the subject of subsequent work \cite{IB+PM+TO=N2}.

In this contribution we describe the results and approach of \cite{Bandos:2023zbs} with emphasis on the superspace formalism and the use of the rheonomic approach to supergravity, and present some initial stages of its generalization to the case of maximal D=11 supergravity, which will be the subject of \cite{IB+PM+TO=D11}.

Our approach to the construction of Noether-Wald charges \cite{Lee:1990nz,Wald:1993nt,Iyer:1994ys} in spacetime supergravity formulation is close to the one in \cite{Barnich:2001jy,Barnich:2003xg}.  To
the best of our knowledge, the concept of Killing supervector was introduced in the book by Buchbinder and Kuzenko \cite{Buchbinder:1995uq} and further developed in \cite{Kuzenko:2012vd,Kuzenko:2015lca,Howe:2015bdd,Howe:2018lwu,Kuzenko:2019tys,Chandia:2022uyy}.

\section{Killing vectors in general relativity and differential forms}

In General Relativity (GR) a Killing vector $k_\mu (x)$ parametrizes a diffeomorphism that leaves a metric invariant, $ - \delta_k g_{\mu\nu}= \nabla_\mu k_\nu + \nabla_\mu k_\mu=: 2 \nabla_{(\mu}
k_{\nu )}=0$.  Hence the standard form of the Killing equation is
\be
\nabla_{(\mu} k_{\nu )}:= \frac 1 2 (\nabla_{\mu} k_{\nu }+\nabla_{\nu} k_{\mu
})=0\; .
\ee
In the tetrad (vielbein) formalism with
\be
\nonumber
g_{\mu\nu}=e_\mu^a\eta_{ab}e_\nu^b \; , \qquad \eta_{ab}=diag (1,-1,-1,-1)\,,
\ee
this corresponds to the statement that the vielbein 1-form $e^a=dx^\mu e_\mu^a$ is invariant {\it up to} SO(1,3) local Lorentz transformations,
\be
\label{vea=Lkea}
- \delta_k e^a := {\cal L}_k e^a := \imath_kde^a+dk^a=
e^bL_{(k)b}{}^a \; , \qquad L_{(k)}^{ab}=- L_{(k)}^{ba}= L_{(k)}^{[ab]} \; ,
\ee where \be de^a= dx^\mu \wedge dx^\nu \partial_{[\nu}e_{\mu ]}^a \; ,
\qquad \imath_k de^a=dx^\mu k^\nu \partial_{[\nu}e_{\mu ]}^a\; , \qquad \imath_k
e^a=k^a=k^\mu e_{\mu }^a \ ,
\ee
and $\wedge$ is the exterior product symbol,
$ dx^\mu \wedge dx^\nu =-dx^\nu \wedge dx^\mu $.

Thus the Killing equation reads $dk^a=-\imath_k(de^a)+e^bL_{(k)b}{}^a$ or
\be
\label{dka=}
        d k^a = - dx^\mu  k^\nu  \partial_{[\nu}e_{\mu ]}^a +
        e^bL_{(k)b}{}^a\; .
\ee
But it is much more convenient to write these equations in a manifestly
covariant form using the spin connection
\be
\label{omega}
    \omega^{ab}=dx^\mu\omega_\mu^{ab}=e^c\omega_c{}^{ab}=-\omega^{ba} , \ee
    the covariant derivative ${\cal D}=dx^\mu {\cal D}_\mu=e^b {\cal D}_b$ and the torsion
      \be {\cal D}e^a=de^a-e^b\wedge \omega_b{}^a= T^a =\frac 1 2 e^c\wedge
      e^b T_{bc}{}^a.
      \ee
      The Killing equation \eqref{dka=} is then written as
       \be\label{Dka=}
 {\cal D}k^a +  \imath_kT^a =  e^b P_{(k)b}{}^a \; ,  \qquad  P_{(k)}{}^{ab}= -\imath_k\omega{}^{ab}+L_{(k)}{}^{ab}
   \; ,
   \ee
    where we have introduced the {\it momentum map}
    $P_{(k)}{}^{ab}= -\imath_k\omega{}^{ab}+L_{(k)}{}^{ba}=- P_{(k)}{}^{ba}$ and used
 \bea\nonumber \imath_k\omega{}^{ab}=k^\mu \omega_\mu{}^{ab}\; , \qquad
    \imath_kT^a =e^ck^bT_{bc}{}^a\, , \qquad  {\rm and} \qquad   {\cal D}k^a= e^b
    {\cal D}_bk^a\, ;
    \eea
when $T^a=0$, all the contributions in the expression for ${\cal D}_bk^a$ become antisymmetric and  ${\cal D}_{(b}k_{a)}=0$ follows, i.e.
\be\nonumber T^{a}=0 \qquad \implies \qquad {\cal D}_{(b}k_{a)}=0
  \; . \ee
This is equivalent to $\nabla_{(\mu}k_{\nu)}=0$ as the spin connection and the affine connection obey the vielbein postulate, that implies the metricity condition for the affine connection when $\omega^{ab}$ preserves $\eta_{ab}$,
    \bea
    \nabla_\mu e_\nu^a &=& \partial_\mu e_\nu^a + \Gamma_{\mu\nu}{}^\rho e_\rho^a -  e_\nu^b\omega_{\mu b}{}^a = 0\;  \qquad \implies \qquad \nabla_\mu  g_{\nu\rho}=0\; .
    \eea

\section{Killing vectors and conserved charges in supergravity}

In the case of supergravity (SUGRA), the set of gauge symmetries includes also supersymmetry (SUSY) transformations which must be taken into account when defining the Killing vector. Furthermore, the
presence of fermionic fields (gravitini) results in a modification of the
Killing equation.  Such a modification can be obtained within the spacetime
component approach to supergravity, but the SUSY transformation properties of
both the Killing vector and the Killing equations do not follow naturally in its frame.

In our \cite{Bandos:2023zbs} we used the superfield approach to ${\cal N}=1$ $D=4$ supergravity to determine these properties and used the Killing vector and generalized Killing spinors thus defined to construct a new Noether-Wald charge and Komar charge of SUGRA including the usually ignored fermionic
contributions.

The first-order action for SUGRA (which also can be used in the 1.5 order formalism) is
\begin{eqnarray}
\label{bfL=4SG}
& S=\int_{M^4} {\mathcal L}_4(e^a,\psi^\alpha,\bar{\psi}{}^{\dot{\alpha}},\omega^{ab}) \; , \qquad {\mathcal L}_4 =\frac {1}{2} \epsilon_{abcd}{\cal R}^{ab}\wedge e^c \wedge e^d +
4 {\cal  D}\psi\wedge \sigma^{(1)}\wedge \bar{\psi}- 4 \psi\wedge \sigma^{(1)}\wedge {\cal D}\bar{\psi},
\end{eqnarray}
where for simplicity we set $16\pi G_N^{(4)}=1$ in intermediate calculations and denoted
\begin{eqnarray}
 & \sigma^{(1)}:= ( e^a\sigma_{a\alpha\dot\alpha}), \qquad  \psi^\alpha =dx^\mu\psi_\mu^\alpha=(\bar{\psi}{}^{\dot\alpha})^*, \qquad  {\cal R}^{ab}=(d\omega-\omega\wedge\omega)^{ab}=\frac 1 2 e^d\wedge e^c  {\cal R}_{cd}{}^{ab} \;  . \qquad
\end{eqnarray}

The action is invariant under three gauge symmetries: (local) Lorentz SO(1,3), diffeomorphisms, and supersymmetry.
 The three Noether currents for these gauge symmetries are trivially
conserved which is to say  their dual 3-forms $ {\cal J}$ are exact,
    \bea \label{eq:Qdef}  & {\cal J}=\frac 1 {3!} e^c\wedge e^b\wedge e^a \epsilon_{abcd}{\cal J}^d = d{\cal Q}\qquad \implies \qquad d{\cal J}\equiv 0 \; \eea
off-shell (let us recall that  the exterior derivative $d=dx^\mu \partial_\mu$ is nilpotent, $dd=0$).
Then, by virtue of Stokes theorem, the conserved charges obtained by
integrating them over closed 3-dimensional surfaces vanish identically and one
must instead use the so-called Noether charges $\mathcal{Q}$ and analyze the
defining equation (\ref{eq:Qdef}) searching for conditions under which
$\mathcal{J}=0$, so that $d\mathcal{Q}=0$.  In simple situations, the Noether
current vanishes identically on-shell for \textit{Killing} (or
\textit{reducibility}) gauge parameters that leave invariant all the fields of
the theory.

\subsection{Conserved charge for local Lorentz symmetry}
In particular, for {\it local Lorentz transformations} given by
\bea
&\delta_L e^a=  e^bL_b{}^a\; , \qquad \delta_L \psi^\alpha = \frac 1 4 \psi^\beta\sigma_{ab\beta}{}^\alpha\, L^{ab}\; , \qquad \delta_L \omega^{ab}={\cal D}L^{ab}\; , \eea
we find, after using the torsion constraint
\be T^a={\cal D}e^a =-2i \psi \wedge \sigma^a\bar{\psi} \; ,\ee
which in the $1^{st}$ order formalism follows from the spin connection's equation of motion, that the corresponding Noether-Wald charge reads
\bea
 & {\cal Q} (L)= - \frac 1 2 L^{ab} \, \epsilon_{abcd}e^c\wedge e^d\; .
\eea

If we consider a $L^{ab}={\mathfrak k}^{ab}$ such that {\it for a solution under consideration} $\delta_{{\mathfrak k}} \psi^\alpha=0$, $\delta_{{\mathfrak k}}\omega^{ab}=0$ (in this case it is not necessary to demand $\delta_{{\mathfrak k}} e^a= 0$), then \be d {\mathfrak{Q}}({\mathfrak  k})\doteq 0  \ee where $\dot{=}$ denotes on-shell equality.
The integral of $ {\mathfrak{Q}}({\mathfrak k})$ over a closed 2-surface $\Sigma^2$ is then the conserved charge
\bea
& {\mathfrak{Q}}({\mathfrak k})=\int_{\Sigma^2} {\cal Q} ({\mathfrak k})= - \frac 1 {32\pi G^{(4)}_N}
\int_{\Sigma^2} {\mathfrak k}^{ab} \, \epsilon_{abcd}e^c\wedge e^d\; .
\eea

\subsection{Conserved supercharge for local supersymmetry}

For local supersymmetry
\begin{align}
\label{susy=}
 \delta_\epsilon e^a & =-2i\psi\sigma^a\bar{\epsilon}+2i\epsilon\sigma^a\bar{\psi}\;
\; , \qquad \delta_\epsilon \psi ={\cal D}\epsilon\; , \qquad \delta_\epsilon\bar{\psi}= {\cal D}\bar{\epsilon}\; , \qquad
\\ \delta_\epsilon \omega^{ab} & \doteq 2i{\cal D}^{[a}\psi^{b]\alpha} (\sigma^{(1)}\bar{\epsilon})_\alpha-  2i(\epsilon\sigma^{(1)})_{\dot\alpha}{\cal D}^{[a}\bar{\psi}{}^{b]\dot\alpha}
\; ,
\end{align}
we find
\bea
 & {\cal Q} (\epsilon )= -4\epsilon \sigma^{(1)}\wedge \bar{\psi} +4\psi\wedge \sigma^{(1)} \bar{\epsilon}
 \; . \qquad
\eea

Choosing an $\epsilon^\alpha=\kappa_s^\alpha$ such that for a solution under consideration
$\delta_{\kappa_s} e^a= 0$ and $\delta_{\kappa_s} \psi^\alpha=0 $, then $d {\frak{Q}}(\kappa_s)\dot{=}0$ and
 \bea
 & {\frak{Q}}(\kappa_s)= \int_{\Sigma^2} {\cal Q} (\kappa_s)=  \frac 1 {4\pi G^{(4)}_N} \int_{\Sigma^2} (-\epsilon \sigma^{(1)}\wedge \bar{\psi} +\psi\wedge \sigma^{(1)} \bar{\epsilon} )\; ,\qquad
\eea
is a conserved supercharge for the supersymmetric solution.

\subsection{On diffeomorphism charge and Killing vectors in supergravity }

The Noether charge associated with diffeomorphisms (usually called Noether--Wald charge) has to be computed taking carefully into account the ``compensating'' Lorentz and SUSY transformations induced by diffeomorphisms.  This is necessary because the invariance of the fields
can only be defined modulo gauge symmetries of the model.
The existence of a superpartner $\kappa^{\alpha}$, called the generalized Killing spinor, of the Killing vector $k^{a}$ is perhaps mysterious in the component approach to supergravity,
but is natural and manifest in the superspace approach (SSP) as they are parts of the so-called Killing supervector, which will be denoted by $K^{A}$;
furthermore, the SSP allows for a rapid deduction of the needed SUSY transformations, a deduction that is more involved in the component approach.
The generalized Killing spinor $\kappa$ enters, as we will see in a few lines, the SUSY transformation of the Killing vector as
    \be \delta_\epsilon k^a = 2i\epsilon\sigma^a\bar{\kappa} -  2i\kappa\sigma^a\bar{\epsilon}
    \; , \qquad \ee
and also in the generic Killing equation of supergravity
\be\label{eq:ESTE}
  {\cal D}k^a- 2i\psi\sigma^a\bar{\kappa}+2i{\kappa}\sigma^a\bar{\psi}=e^b P_{(K)b}{}^a\; .  \qquad
\ee

Moreover, the SSP gives {\it generalized Killing spinor equations} for $\kappa$, and determines the supersymmetry properties of the Killing  vector equation (\ref{eq:ESTE}).

\section{Killing supervectors in simple $\mathcal{N}=1$ $D=4$ supergravity superspace}

In curved superspace with coordinates
\begin{equation}\nonumber
Z^M= (x^\mu, \theta^{\check{\underline{\alpha}}}) \;  \qquad \mu=0,1,2,3\; , \qquad {\check{\underline{\alpha}}}=1,2,3,4\;
\end{equation}
simple  ${\cal N}=1$ D=4 supergravity is described by the supervielbein and the spin connection 1-forms
\begin{equation}\label{EA=4D1N}
E^A=dZ^ME_M^A(Z)=(E^a,E^{\underline{\alpha}})=(E^a,E^{\alpha},{\bar E}{}^{\dot{\alpha}})\; ,\qquad \omega^{ab}=dZ^M\omega_M^{ab}(Z)=E^C\omega_C^{ab}=-\omega^{ba}\; ,
\end{equation}
where we split the spinor index $\underline{\alpha}=1,2,3,4$ into dotted ($\dot{\alpha}=1,2$) and undotted ($\alpha =1,2$) indices, corresponding to
Weyl spinors. The supervielbein and the connection obey a set of constraints that are imposed on the torsion and the curvature, whose definition in the SSP reads
\begin{align}
 T^A & = DE^A = dE^A-E^B\wedge \omega_B{}^A= \frac 1 2 E^C\wedge E^B T_{BC}{}^A\; , \qquad
 \\
 R^{ab} & =d\omega^{ab}-\omega^{ac}\wedge \omega_c{}^{b}=\frac 1 2 E^D\wedge E^C R_{CD}{}^{ab}\;  ,
\\
 & {\rm where} \qquad  \omega_B{}^C ={\rm diag}(\omega_b{}^c, \omega_\beta{}^\gamma, \omega_{\dot{\beta}}{}^{\dot{\gamma}} )\; , \qquad \omega_\beta{}^\gamma =\frac 1 4  \omega^{ab}\sigma_{ab \beta}{}^\gamma\, , \qquad \omega_{\dot{\beta}}{}^{\dot{\gamma}} =-\frac 1 4 \omega^{ab}\tilde {\sigma}_{ab}{}^{\dot{\gamma}}{}_{\dot{\beta}}\; .
\end{align}
We will detail the constraints later on, but want to stress that local spacetime supersymmetry of supergravity 
comes from the superdiffeomorphism invariance of the superspace formalism.

Notice that wedge product of fermionic superforms is symmetric, e.g. $E^\alpha \wedge E^\beta = + E^\beta \wedge E^\alpha$, whereas  $E^\alpha \wedge E^b = - E^b \wedge E^\alpha$ and
${\bar E}{}^{\dot\alpha} \wedge E^b = - E^b \wedge {\bar E}{}^{\dot\alpha}$.

A {\it Killing supervector} \cite{Buchbinder:1995uq} in curved ${\cal N}=1$ superspace
\be  K^A=(K^a, K^{\underline{\alpha}})=(K^a, K^\alpha, \bar{K}{}^{\dot{\alpha}})\; , \qquad a=0,1,2,3\; , \qquad \underline{\alpha}=1,..., 4 \ee
is defined by the conditions (valid in this form for a generic superspace)
 \begin{align}\label{DKA=sKilling}
 -\delta_K E^A & = DK^A + \imath_K T^A + E^B \imath_K{\omega}_B{}^A= E^B {L}_{(K)B}{}^A \; , \qquad \\ \label{iKR=sKilling}
 -\delta_K \omega^{ab} & = D\imath_K\omega^{ab} + \imath_K R^{ab} = D {L}_{(K)}^{ab}\; , \qquad
 \end{align}
where
\be D= E^AD_A=E^aD_a+E^{\underline{\alpha}}D_{\underline{\alpha}}=dZ^MD_M\; , \qquad
 \imath_K \omega^{ab}=K^C\omega_C^{ab}(Z), \qquad \imath_K T^A= E^CK^B T_{BC}^{A}(Z)\ee
and ${L}_{(K)B}{}^A$ denotes a compensating local Lorentz transformation. In the specific case of ${\cal N}=1$, $D=4$ supergravity we have
\be  L_{(K)B}{}^C
 ={\rm diag}(L_b{}^c, L_\beta{}^\gamma, L_{\dot{\beta}}{}^{\dot{\gamma}} )\; , \qquad L_\beta{}^\gamma =\frac 1 4  L^{ab}\sigma_{ab \beta}{}^\gamma\, , \qquad L_{\dot{\beta}}{}^{\dot{\gamma}} =-\frac 1 4 L^{ab}\tilde {\sigma}_{ab}{}^{\dot{\gamma}}{}_{\dot{\beta}}
 \; . \ee

It is convenient to define the following momentum map superfield
\bea\label{Pab:=}
P_{(K)B}{}^A = -\imath_K{\omega}_B{}^A+ {L}_{(K)B}{}^A = {\rm diag}(P_{(K)b}{}^a, P_{(K)\beta}{}^\gamma, P_{(K)\dot{\beta}}{}^{\dot{\gamma}} )\; , \qquad \;  \eea
and write the above {\it {superKilling equations}} \eqref{DKA=sKilling} and \eqref{iKR=sKilling} as
\bea\label{DKA=sK}
&& \fbox{$DK^A + \imath_K T^A = E^B P_{(K)B}{}^{A}$}\; , \qquad \\    \label{iKR=sK}
&& \fbox{$ DP_{(K)}{}^{ab} = \imath_K R^{ab}  \equiv E^DK^CR_{CD}{}^{ab} $}  \; . \qquad
\eea

\subsection{Constraints of simple supergravity, its superspace torsion and curvature }

The curved superspace of simple ${\cal N}=1$, $D=4$ on-shell supergravity
is defined by the torsion constraints which result in
\bea\label{TA=on-shell}
T^a&=& DE^a= -2i E^\alpha\wedge \bar{E}{}^{\dot{\alpha}} \sigma^a_{\alpha\dot{\alpha}}
\; , \qquad
T^{\alpha}= DE^{\alpha}
=
\frac
1{2} E^{c}\wedge E^{b} T_{bc}{}^{\alpha}\; , \qquad
\\   R^{ab} & =&-2i \sigma^{(1)}_{\alpha\dot{\beta}}
\wedge E^{\dot{\beta}} T^{ab\alpha}+2i E^\beta \wedge \sigma^{(1)}_{\beta\dot{\alpha}}
 T^{ab\dot\alpha}+ \frac 1 2 E^d\wedge E^c R_{cd}{}^{ab}\; , \nonumber
\eea
where
\be\label{D=EADA} D= E^AD_A=E^aD_a+E^{\alpha}D_{\alpha}+E^{\dot{\alpha}}D_{\dot{\alpha}}\; .
\qquad
\ee
Moreover, in \eqref{TA=on-shell} the superfield generalization of the gravitino field strength, $T_{bc}{}^{\alpha}(Z)$,
obeys the superfield generalization of the Rarita-Schwinger equation
\be\label{RS=SSP}
 \epsilon^{abcd}T_{ab}{}^{\alpha}\sigma_{c\alpha\dot\beta}=0\qquad \Rightarrow \qquad \begin{cases} T_{ab}{}^{\alpha}= \frac i 2 \epsilon_{abcd} T{}^{cd\alpha} \; , \cr T_{ab}{}^{\dot\alpha} =- \frac i 2 \epsilon_{abcd}   T^{cd\dot\alpha} \, . \qquad \end{cases}
\ee
and
$R_{cd}{}^{ab}(Z)$ obeys the superfield generalization of the Einstein equation
\begin{eqnarray}\label{EinEqZ=N2}
R_{ab}{}^{cb}(Z)-\frac 12 \delta_{a}{}^{c}R_{ef}{}^{ef}(Z) = 0.   \qquad
\end{eqnarray}

\subsection{Generalized action of the Rheonomic approach to supergravity.}

\label{GAPsection}
Where do the above constraints comes from?
One way to obtain them is
from the generalized action principle of the rheonomic approach to supergravity \cite{Neeman:1978njh,DAuria:1982uck,Castellani:1991eu}
which can be obtained from the $1^{st}$ order action \eqref{bfL=4SG} by substituting
\begin{eqnarray}
\label{e->E}
e^a(x)\mapsto E^a(x,\theta) , \quad \psi^\alpha (x)\mapsto E^\alpha(x,\theta),\quad \omega^{ab}(x)\mapsto
dZ^M\omega_M^{ab}(Z) , \quad {\cal D}=dx^\mu{\cal D}_\mu \mapsto D=dZ^MD_M
\end{eqnarray}
and replacing the integration over spacetime by an integration over an arbitrary surface of maximal bosonic dimension in superspace, determined by fermionic coordinate functions
$\theta (x)$,
\be\nonumber
{\cal M}^4\in \Sigma^{(4|4)}\; : \qquad \theta =\theta (x) , \qquad x=\text{arbitrary} \; .
\ee

In short, the rheonomic approach allows to lift the equations of motion obtained from the $1^{st}$ order action \eqref{bfL=4SG} to the superspace equations by \eqref{e->E}.
The aforementioned  constraints and superspace equations of motions can be obtained in
this manner,
and allows us to establish the direct relation between the results and calculations in
the SSP and the spactime component approach to SUGRA. Actually, one more ingredient for this relation to hold is necessary: the Wess-Zumino gauge (WZ).

\subsection{Wess-Zumino gauge}

To pass to the spacetime component formulation of SUGRA, besides imposing the constraints,
we should use the superdiffeomorphism symmetry and superspace local Lorentz symmetry to fix the WZ gauge
$ \theta^{\check{\underline{\alpha}}} E_{\check{\underline{\alpha}}}{}^A  = \theta^{\check{\underline{\alpha}}}  \delta_{\check{\underline{\alpha}}}{}^A$, $  \theta^{\check{\underline{\alpha}}} w_{\check{\underline{\alpha}}}{}^{ab}=0$
or, equivalently
\begin{eqnarray}\label{WZg}  \imath_{\underline{\theta}} E^{a}=0\; , \qquad  \imath_{\underline{\theta}} E^{\underline{\alpha}}=
\theta^{\underline{\alpha}} \; , \qquad  \imath_{\underline{\theta}} w^{ab}=0\; . \end{eqnarray}
This gauge is invariant under spacetime diffeomorphisms, spacetime local Lorentz symmetry and local spacetime supersymmetry only.

In the WZ gauge we have
  \begin{eqnarray} \label{WZ0gg1} & E_N{}^A\vert_{\theta =0} = \left(\begin{matrix}e_\nu^a(x) &
\psi_\nu{}^{\underline{\alpha}}(x)\cr 0 & \delta_{\check{\underline{\beta}}}{}^{\underline{\alpha}} \end{matrix}\right) \; , \qquad
\label{WZ0gg2} & E_A{}^N\vert_{\theta =0} =
\left(\begin{matrix}e_a^\nu(x) & - \psi_a{}^{\check{\underline{\beta}}}(x)\cr 0 &
\delta_{\underline{\alpha}}{}^{\check{\underline{\beta}}}\end{matrix}\right) \; ,  \end{eqnarray}
$\theta^{\underline{\beta}} \; {\cal
D}_{\underline{\beta}}   =: \theta {\cal D} =\theta\partial
 :=  \theta^{\check{\underline{\alpha}}} \partial_{\check{\underline{\alpha}}}
$, but
$(D_{a}(...))\vert_{\theta=0} = (E_{a}^M \partial_M (...))\vert_{\theta=0} =  e_a^\mu \partial_\mu  ((...)\vert ) -\psi_a{}^{{\underline{\alpha}}} (D_{\underline{\alpha}}(...))\vert  $
and
\begin{eqnarray}\nonumber
\label{Tbbf0WZ=Gen} T_{ab}{}^{\underline{\alpha}}\vert_{\theta=0}  &=& e_a^\mu e_b^\nu T_{\mu\nu}{}^{\underline{\alpha}}(x) - 2\psi_{[a|}{}^{\underline{\beta}} T_{\underline{\beta}|b]}{}^{\underline{\alpha}}\vert_0 -  \psi_{b}{}^{\underline{\beta}} \psi_{a}{}^{\underline{\gamma}} T_{\underline{\gamma}\underline{\beta}}{}^{\underline{\alpha}}\vert_{\theta=0}
 \; , \\ \nonumber
\label{Tbbb0WZ=Gen} T_{ab}{}^{c}\vert_{\theta=0}  &=& e_a^\mu e_b^\nu T_{\mu\nu}{}^{c}(x) - 2\psi_{[a|}{}^{\underline{\beta}} T_{\underline{\beta}|b]}{}^{c}\vert_{\theta=0} -  \psi_{b}{}^{\underline{\beta}} \psi_{a}{}^{\underline{\gamma}} T_{\underline{\gamma}\underline{\beta}}{}^{c}\vert_{\theta=0}
 \; , \\ \nonumber
 \label{Rcdab0WZ=Gen}
R_{cd}{}^{ab}\vert_{\theta=0} &=& e_c^\mu e_d^\nu R_{\mu\nu}{}^{ab}(x) - 2\psi_{[c|}{}^{\underline{\alpha}} R_{\underline{\alpha}|d]}{}^{ab}\vert_{\theta=0} - \psi_{d}{}^{\underline{\beta}} \psi_{c}{}^{\underline{\alpha}} R_{\underline{\alpha}\underline{\beta}}{}^{ab}\vert_{\theta=0}\; .  \qquad
  \end{eqnarray}
We want to stress
that the curved superspace index $\check{\underline{\alpha}}$ of the fermionic coordinates of curved superspace can be identified as a spinor index $\underline{\alpha}$,
only after this gauge fixing.

\subsection{Killing supervector equations and Killing equation of simple Poincaré supergravity}

The superspace superKilling equations \eqref{DKA=sK}
in the case of simple $\mathcal{N}=1$ $D=4$ supergravity splits into
\bea\label{DKa=}
DK^{a}&=& 2i E^{\alpha}\sigma^a_{\alpha\dot{\alpha}}K^{\dot{\alpha}}
 +2i  \bar{E}_{\dot{\alpha}}\tilde{\sigma}^{a\dot{\alpha}\alpha} K_{\alpha}+
 E^{b} P_{(K)b}{}^a\; , \qquad    \qquad \\
\label{DKal=}
DK^\alpha &=&
- E^bK^aT_{ab}{}^\alpha + E^\beta P{}_{(K)\beta}{}^\alpha \, , \qquad
P_{(K)\beta}{}^{\alpha} =-\frac 1 4 P_{(K)}^{ab}{\sigma}{}_{ab}{}_{\beta}{}^{\alpha}  \; ,  \qquad \eea
and the c.c. of the latter.
Using \eqref{D=EADA},
we find that the first of these equations splits into
\be\label{DbDKa==}  D_b K^a=P_{(K) b}{}^a \qquad \implies \qquad
D^{(a}K^{b)}=0\; , \qquad P_{(K)}^{ab}=D^{[a}K^{b]}\,,
\ee
which gives the superfield generalization of the standard Killing equation and
of the definition of the momentum map, and
\be\label{DfKa==}
D_\alpha K^a= 2i \sigma^a_{\alpha\dot{\alpha}} K^{\dot{\alpha}} \; , \qquad
\bar{D}_{\dot\alpha} K^a= 2i K^{\alpha}\sigma^a_{\alpha\dot{\alpha}}\;,
\ee
which determines the supersymmetry transformations of the Killing vector in
supergravity.

Indeed, denoting the leading components of the bosonic Killing supervector superfield by $k^a(x)$ and $\kappa^\alpha$,
\be K^a\vert_{\theta=0}=k^a(x), \qquad K^\alpha\vert_{\theta=0}=\kappa^\alpha \; , \ee
we find that in the WZ gauge \eqref{WZg}, where \eqref{WZ0gg1} holds,
the leading component of Eq. \eqref{DbDKa==} can be written as
\bea\label{cDka=}
\fbox{${\cal D}k^a- 2i\psi\sigma^a\bar{\kappa}+2i{\kappa}\sigma^a\bar{\psi}=e^b P_{(K)b}{}^a$}\; \quad
\implies \quad \begin{cases} {\cal D}^{(a}k^{b)}- 2i\psi^{(a}\sigma^{b)}\bar{\kappa}+c.c.=0 \; , \cr P_{(K)}^{ab}(x)={\cal D}^{[a}k^{b]}- 2i\psi^{[a}\sigma^{b]}\bar{\kappa}+c.c. \end{cases} \eea
This gives the {\it complete set of bosonic Killing equations} for supergravity.

The leading component of Eq.~\eqref{DfKa==} determine, through the Lie derivative representation of superdiffeomorphisms, the {\it SUSY transformation of the Killing vector}, namely
\be\nonumber \fbox{$ \delta_\epsilon k^a(x)=  2i \epsilon\sigma^a\bar{\kappa} - 2i \kappa\sigma^a\bar{\epsilon}$} \; . \ee

 The second superKilling equation
\eqref{DKal=}
splits into
\bea\label{DbKal=}
 D_bK^\alpha &=&
- K^aT_{ab}{}^\alpha \; , \qquad \\ \label{DfKal=}  D_\beta K^\alpha & =& P{}_{(K)\beta}{}^\alpha:= \frac 1 4 P{}_{(K)}^{ab}\sigma_{ab\beta}{}^\alpha \; , \qquad \bar{D}_{\dot\beta } K^\alpha  =  0\, . \qquad
\eea
 The leading component of Eq. \eqref{DbKal=} (in the WZ gauge) leads to the {\it generalized Killing spinor equation}
\bea
\label{cDkap=} & \fbox{$ {\cal D}\kappa^\alpha= -\imath_k ( {\cal D}\psi^\alpha)+\psi^\beta P_{(K)\beta}{}^\alpha$}\, , \qquad  \eea
where $\imath_k ( {\cal D}\psi^\alpha)= 2e^ck^b  {\cal D}_{[b}\psi_{c]}^\alpha $, $P_{(K)\beta}{}^\alpha=\frac 1 4 P_{(K)}^{ab}{\sigma}{}_{ab}{}_{\beta}{}^{\alpha}$,
while the leading component of Eq. \eqref{DfKal=} encodes the SUSY transformation of the generalized
fermionic Killing spinor
    \bea   & \fbox{$\delta_\epsilon {\kappa}^{\alpha} = \epsilon^{\beta} P_{(K)\beta}{}^\alpha =
\frac 1 4 (\epsilon\sigma_{ab})^\alpha \left({\cal D}^{[a} k^{b]}
+2i {\kappa}^{\gamma}(\sigma^{[a}\bar{\psi}{}^{b]})_{\gamma}+c.c. \right)$} . \qquad\eea

 Similarly, the SSP equation for the momentum map, Eq. \eqref{iKR=sK}, in the case of  simple SUGRA reads
\bea
 \label{DPK=}
DP_{(K)}^{ab}&=&K^c \left(E^dR_{cd}{}^{ab}+2iE_{\dot{\beta}}\tilde{\sigma}_c^{\dot{\beta}\alpha}T^{ab}{}_\alpha
+2iE^{{\beta}}{\sigma}_{c\beta\dot{\alpha}}T^{ab\dot{\alpha}}\right)-
 2iK_{\dot{\beta}}\tilde{\sigma}^{(1)\dot{\beta}\alpha}T^{ab}{}_\alpha
-2iK^{{\beta}}{\sigma}^{(1)}_{\beta\dot{\alpha}}T^{ab\dot{\alpha}}\quad
\eea
 and encodes  the spacetime equation for the momentum map
\bea\label{cDPab=}
{\cal D}P^{ab}_{(k,\kappa)}= \imath_k{\cal R}{}^{ab}-4i ((\kappa -\imath_k\psi) \sigma^{(1)})_{\dot\alpha} {\cal D}^{[a}\bar{\psi}{}^{b]\dot{\alpha}}+ 4i  {\cal D}^{[a}{\psi}{}^{b]{\alpha}}
(\sigma^{(1)}(\bar{\kappa} -\imath_k\bar{\psi}))_{\alpha} \;
\eea
as well as its supersymmetry transformations
\bea\label{susyPK=}
\delta_\epsilon P_{(k,\kappa)}^{ab}= 4ik^c \left((\epsilon{\sigma}_{c})_{\dot{\alpha}}{\cal D}^{[a}\bar{\psi}{}^{b]\dot{\alpha}}+
 (\bar{\epsilon} \tilde{\sigma}_{c})^{\alpha}{\cal D}^{[a}{\psi}{}^{b]}{}_{\alpha}\right). \qquad
\eea

Using these results we can show that the Killing equation \eqref{cDka=} and the generalized Killing spinor equation \eqref{cDkap=} for supergravity form a supermultiplet under supersymmetry transformations \eqref{susy=} \cite{Bandos:2023zbs}.
This is actually guaranteed by the superspace origin of these equations, as was described above.

To be more precise, under supersymmetry  \eqref{cDka=} is transformed by \eqref{cDkap=}, while supersymmetry transformations of \eqref{cDkap=} is expressed in terms of  \eqref{cDPab=}. However, as far as \eqref{cDPab=} can be obtained as consistency condition of  \eqref{cDkap=}, we have the usual situation when the boson(ic equation) is transformed through fermion(ic equation), while fermion(ic equation) transforms through the derivative of the boson(ic equation).

\section{Noether-Wald charge and Komar charge of simple supergravity}

The Noether-Wald charge is the 2-form associated to the invariance under diffeomorphisms. The conserved charge, i.e. an on-shell closed 2-form, should be associated with diffeomorphisms parametrized by Killing vectors. But in supergravity the diffeomorphism generated by a Killing vector appears accompanied by compensating supersymmetry and local Lorentz transformations.  Thus to construct the correct Noether-Wald charge, we have to consider diffeomorphisms $\delta_\xi$ accompanied by (induced or associated) local Lorentz and local SUSY transformations, $\delta_{L_\xi}$ and $\delta_{\epsilon_\xi}$.

Actually, only these two give a contribution to the total derivative term in the expression for on-shell variations of the Lagrangian,
i.e. in the conserved current which is therefore expressed in terms of the
momentum map $P_{\xi}^{ab}=-\imath_\xi \omega^{ab}+L_\xi^{ab} $ and the fermionic ${\epsilon_\xi}$.
The result obtained in \cite{Bandos:2023zbs} is
 \begin{eqnarray}
  \label{eq:Noether-Waldcurrent}
  \mathbf{J}[\xi,\epsilon_\xi]
  &=&  -\frac 1 2  \epsilon_{abcd} e^{c}\wedge e^{d} \wedge \mathcal{D}P_{\xi}{}^{ab} + 2i \epsilon^{abcd} P_{\xi}{}^{ab} e_{c}\wedge \psi\sigma^d\wedge \bar{\psi} - \nonumber \\
  && -4\epsilon_\xi \sigma^{(1)}\wedge\mathcal{D} \psi  + 4 \mathcal{D} \psi \sigma^{(1)}\wedge \bar{\epsilon}_\xi
  -4\mathcal{D}\epsilon_\xi \sigma^{(1)}\wedge \psi  + 4 \psi \sigma^{(1)}\wedge \mathcal{D}\bar{\epsilon}_\xi \,,
\end{eqnarray}

As expected, this Noether current 3-form is exact
   \begin{eqnarray}
  \label{eq:dJ=Qdiff}
  {\mathbf{J}}[\xi,\epsilon_\xi] =
  d{\mathbf{Q}}[\xi,\epsilon_\xi]\,, \\  \label{eq:Qdiff=}
  {\mathbf{Q}}[\xi,\epsilon_\xi]
  = -\frac 1 2 \epsilon_{abcd}P_k{}^{ab}e^c\wedge e^d-4 {\kappa}\sigma^{(1)}\wedge \bar{\psi} +4\psi\wedge \sigma^{(1)} \bar{\kappa}\quad
\end{eqnarray}
and its 2-form potential, $\mathbf{Q}[\xi,\epsilon_\xi]$ in \eqref{eq:Qdiff=},
is the {\it Noether-Wald charge} 2-form we were after. Indeed, it is
manifestly invariant under diffs and local $SO(1,3)$.

Furthermore, for Killing parameters $(k,\kappa)$ this Noether-Wald charge 2-form is on-shell closed and supersymmetry-invariant up to a
total derivative, i.e.
\begin{eqnarray}
    & d\mathbf{Q}[k,\kappa]
     \doteq
      0\,,  \qquad  \delta_{\epsilon}\mathbf{Q}[k,\kappa]
 \doteq
-4 d\left( \epsilon\sigma^{(1)}\bar{\kappa} + \kappa\sigma^{(1)}\bar{\epsilon} \right)\,.
  \end{eqnarray}
Thus $\mathbf{Q}[k,\kappa]$ is {\it SUSY generalization of the
Komar charge} 2-form.  Observe that it is the sum of a term corresponding to the
standard gravitational Komar charge ($\propto$ the momentum map $P_{\xi\, ab}$) and a term
corresponding to the supercharge ($\propto {\kappa}$ and $\propto \bar{\kappa}$),
neither of which being invariant under SUSY.

The last observation is that the above properties of the Komar charge 2-form actually follows from the fact that it is equal to -and can be obtained as- the 'body part' of a closed 2-form ${\mathbf{Q}}[K(Z)]$ in the on-shell SUGRA superspace, i.e.  $ {\mathbf{Q}}[\xi,\epsilon_\xi]={\mathbf{Q}}[K(Z)]\vert_{\theta=0}$. In its turn, ${\mathbf{Q}}[K(Z)]$ obeying $d{\mathbf{Q}}[K(Z)]=0$  is obtained
by lifting to superspace of the spacetime 2-form \eqref{eq:Qdiff=} and reads
 \begin{eqnarray}
 \label{eq:Qdiff=SSP}
  {\mathbf{Q}}[K(Z)]
  = -\frac 1 2 \epsilon_{abcd}P_{(K)}{}^{ab}E^c\wedge E^d-4 K^\alpha \sigma^{(1)}_{\alpha\dot{\alpha}}\wedge E^{\alpha\dot{\alpha}} +4 E^{\alpha}\wedge \sigma^{(1)}_{\alpha\dot{\alpha}} \bar{K}{}^{\dot{\alpha}}\; , \qquad \\  d{\mathbf{Q}}[K(Z)]=0\; , \qquad
   {\mathbf{Q}}[K(Z)]\vert_{\theta=0}=  {\mathbf{Q}}[\xi,\epsilon_\xi]\; . \qquad
\end{eqnarray}

Thus, the Komar 2-form can be found by searching for a closed 2-form in an
on-shell supergravity superspace with certain 'initial conditions' imposed on
its 'body' or 'leading component', as given by its $\theta=0$ 'value'.  This
observation is optional in the case of simple ${\cal N}=1$, $D=4$ Poincar\'e
supergravity, but becomes very helpful in more complicated cases, beginning
from minimal ${\cal N}=2$, $D=4$ Poincar\'e supergravity, which
will be the subject of \cite{IB+PM+TO=N2}.  Here,
instead of addressing this case, we will describe the first stages of the
generalization of our  approach to the case of eleven-dimensional supergravity.

\section{Killing supervectors of 11D supergravity }

In this section $\mu,\nu, \rho,\sigma =0,1,...,9,10$ denote the 11-vector world (curved space) indices,  $a,b,c,d =0,1,...,9,10$  are  the tangent space (flat) vector indices
and Greek letters  denote 11D Majorana spinor indices, $\alpha,\beta, \gamma, \delta =1,...,32$.
We use the mostly minus metric convention
for which the 11D Dirac matrices $\Gamma^{a}{}_{\alpha}{}^{\beta}$, obeying
\begin{eqnarray} \Gamma^{a}{}_{\alpha}{}^{\beta}\Gamma^{b}{}_{\beta}{}^\gamma+\Gamma^{b}{}_{\alpha}{}^{\beta}\Gamma^{a}{}_{\beta}{}^\gamma = \eta^{ab}\delta{}_{\alpha}{}^\gamma\; , \qquad \eta^{ab}={\text{diag}} (1,-1,...,-1)\; ,
\end{eqnarray}
are purely imaginary.
The Majorana spinor indices  ${\alpha}, {\beta}, \gamma=1,...,32$ are lowered and raised by the charge conjugation matrix $C_{\alpha\beta}=- C_{\beta\alpha}$, which is imaginary and antisymmetric, and by its inverse $C^{\alpha\beta}=- C^{\beta\alpha}$, respectively. In the equations below we use the real symmetric matrices
\begin{eqnarray}
  \Gamma^{a}{}_{\alpha\beta}= \Gamma^{a}{}_{\alpha}{}^{\gamma}C_{\gamma\beta}
  =
  \Gamma^{a}{}_{\beta\alpha}\;   , \qquad \tilde{\Gamma}^{a \; \alpha\beta}= C^{\alpha\gamma} {\Gamma}^{a \;\beta}_\gamma = \tilde{\Gamma}^{a \; \beta\alpha} \; ,
\end{eqnarray}
the matrix-valued differential forms $\bar{\Gamma}^{(1)}_{\alpha\beta}:= e^a {\Gamma}_{a\, \alpha\beta}$,
\begin{eqnarray}
\label{Gammak:=} \bar{\Gamma}^{(k)}_{\alpha\beta} &:=&\!\! \frac {1}{
k!} e^{a_k} \wedge \ldots \wedge e^{a_1} \Gamma_{a_1 \ldots a_k}
{}_{\alpha\beta}:= \frac{(-1)^{n(n-1)/2}}{ k!}
\bar{\Gamma}^{(1)}{}_\alpha{}^{\beta_1}\wedge
\bar{\Gamma}^{(1)}{}_{\beta_1}{}^{\beta_2}\wedge  \ldots \wedge
\bar{\Gamma}^{(1)}{}_{\beta_{k-1} \beta} \;,
\end{eqnarray}
and their superspace generalizations with $e^a=dx^\mu e_\mu^a(x) \mapsto E^a=dZ^M E_M^a(Z)$.

\subsection{First-order action of 11D supergravity }

The first-order action for 11D  supergravity can be written as an integral
\begin{eqnarray}\label{S11:=}
S=\int_{{M}^{11}}{\cal
L}_{11}[e^a, \psi^\alpha, \omega^{ab}, A_3 ,
F_{a_1a_2a_3a_4}]\,,
\end{eqnarray}
of the 11-form Lagrangian  \cite{DAuria:1982uck,Julia:1999tk,Bandos:2004ym}
\begin{eqnarray}\label{L11:=}
{\cal L}_{11} &=& \frac {1}{4} R^{ab}\wedge e^{\wedge 9}_{ab} - D\psi^\alpha \wedge
\psi^\beta   \wedge \bar{\Gamma}^{(8)}_{\alpha\beta} +   \frac {1}{4}  \psi^\alpha \wedge
\psi^\beta   \wedge \left(T^a + \frac i 2 \, \psi \wedge \psi \, \Gamma^a\right)  \wedge e_a
\wedge \bar{\Gamma}^{(6)}_{\alpha\beta}  + \nonumber \\
&+&
(dA_3- a_4) \wedge (\ast F_4 + b_7) - \frac {1}{2} F_4 \wedge \ast F_4  +  {1\over 2} a_4 \wedge b_7 -
{1\over 3} A_3 \wedge dA_3\wedge dA_3  \; .
\end{eqnarray}
The bosonic and fermionic 1-forms $e^a=dx^\mu e_\mu^a$ and $\psi^\alpha=dx^\mu \psi_\mu^\alpha$ describe the graviton and the gravitino,
and we have defined
\begin{eqnarray}\label{a4:=}
a_4:=  \frac {1}{2}\psi^\alpha \wedge \psi^\beta   \wedge
\bar{\Gamma}^{(2)}_{\alpha\beta} \; , \qquad
\label{b7:=} b_7:=   \frac {i}{2} \psi^\alpha \wedge \psi^\beta   \wedge
\bar{\Gamma}^{(5)}_{\alpha\beta} \; ,
\qquad \\
\label{E11-n:=} e^{\wedge (11-k)}_{a_1\ldots a_k} :=  \frac {1}{(11- k)!}
\varepsilon_{a_1\ldots a_kb_1\ldots b_{11-k}} e^{b_1}\wedge \ldots
e^{b_{11-k}} \; .
\end{eqnarray}
The {\it purely bosonic} forms  $F_4$, $\ast F_4$ are constructed
out of the auxiliary antisymmetric tensor
$F_{abcd}(x)$ as
\begin{eqnarray}\label{F4:=M11}
F_4&:=&  \frac {1}{4!} \; e^{a_4} \wedge \ldots \wedge e^{a_1} F_{a_1\ldots a_4} \; ,
\\\label{*F4:=M11}
\ast F_4&:=&     \frac {1}{7!\, 4!}\; e^{b_7} \wedge \ldots \wedge e^{b_1}
 \; \varepsilon_{b_1\ldots b_7a_1\ldots a_4} F^{a_1\ldots a_4}=-\frac 1 {4!} F^{abcd}e^{\wedge 7}_{abcd}\; . \qquad
\end{eqnarray}
This auxiliary tensor field, as well as the spin connection 1-form $\omega^{ab}=dx^\mu \omega_\mu{}^{ab}=-\omega^{ba}$,
are independent variables of the first-order action \eqref{S11:=}.

The action
\eqref{S11:=} with
\eqref{L11:=} is invariant under the supersymmetry transformations with
\begin{eqnarray}\label{vea=susy11D}
\delta_{\epsilon} e^a &=& - 2 i \psi^\alpha \Gamma^a_{\alpha\beta} \epsilon^\beta  \; ,  \qquad\label{vA3=susy11D}
\delta_{\epsilon} A_3 = \psi^\alpha  \wedge
\bar{\Gamma}^{(2)}_{\alpha\beta} \epsilon^\beta  \; ,
\\ \label{vpsi=susy11D}
\delta_{\epsilon}  \psi^\alpha &=& {\cal D}\epsilon^\alpha := {D}\epsilon^\alpha  - \epsilon^\beta t_{1\beta}{}^{\alpha} \; ,  \qquad
\end{eqnarray}
where we defined
\begin{eqnarray}\label{t1:=}
  && t_{1\beta}{}^{\alpha}
     =
      \frac {i}{18} e^a
\left(F_{ac_1c_2c_3}\Gamma^{c_1c_2c_3} + {\frac 1 8}\;
F^{c_1c_2c_3c_4} \Gamma_{a c_1c_2c_3c_4}\right){}_\beta^{\;\;\alpha}  \; , \qquad
\end{eqnarray}
which have to be
supplemented by certain transformations of the $F_{abcd}$ field and the $\omega^{ab}$ 1-form.

\subsection{Equations of motion }
The equation of motion for the spin connection determines the spacetime torsion
to be
\begin{equation}\label{Ta=spacetime}
T^a= -i\psi\Gamma^a\wedge \psi\;,
\end{equation}
while the equation for the auxiliary antisymmetric tensor gauge field
$F_{abcd}(x)$ relates it to the field strength of the 3-form gauge field by
\begin{eqnarray}\label{cF4=}
{\cal F}_4:=dA_3=a_4 + F_4 =  \frac {1}{2} \psi^\alpha \wedge \psi^\beta   \wedge
\bar{\Gamma}^{(2)}_{\alpha\beta}+  \frac {1}{4!} \; e^{a_4} \wedge \ldots \wedge e^{a_1} F_{a_1\ldots a_4} .
\end{eqnarray}

The variation with respect to the 3-form gauge field $A_3$
results in the dynamical field equation
\begin{eqnarray}\label{d*F4=}d(*F_4- A_3\wedge dA_3+b_7)=0\; . \qquad
\end{eqnarray}
This equation can be represented as the Bianchi identity
\begin{eqnarray}\label{dF7=F4F4} d{\cal F}_7 - {\cal F}_4\wedge {\cal F}_4=0\; . \qquad
\end{eqnarray}
for the 7-form field strength of the dual 6-form gauge field
\begin{eqnarray}\label{cF7=11D}{\cal F}_7=dA_6+ A_3\wedge dA_3 =b_7 + F_7 =  \frac {i}{
2} \psi^\alpha \wedge \psi^\beta   \wedge
\bar{\Gamma}^{(5)}_{\alpha\beta}+  \frac {1}{7!} \; e^{a_7} \wedge \ldots \wedge e^{a_1} F_{a_1\ldots a_7} \; ,
\end{eqnarray}
if we define that seven-th rank antisymmetric tensor $F_{c_1 \ldots c_7}$ to be dual to the 4-th rank $F_{abcd}$ in \eqref{cF4=},
\begin{eqnarray}\label{F=*F}
F_{c_1 \ldots c_7} = (\ast F_4)_{c_1 \ldots c_7}:=  \frac {1}{4!}
\varepsilon_{c_1 \ldots c_7b_1 \ldots b_4} F^{b_1 \ldots b_4} \qquad \Leftrightarrow \qquad F_7=*F_4\; .
\end{eqnarray}

After some algebra, one finds that the gravitino field equation has the relatively simple form of \cite{DAuria:1982uck,Julia:1999tk}
\begin{eqnarray}\label{RS=11D}
{\cal E}_{10\; \alpha} ={\hat{{\cal D}}}\psi^\alpha \wedge \Gamma^{(8)}_{\alpha\beta}=0 \, ,
\end{eqnarray}
where $\hat{{\cal D}} \psi^\alpha$ is a covariant derivative with generalized connection \cite{Duff:2003ec,Hull:2003mf} (see also \cite{Bandos:2005mm} and refs. therein)
\begin{eqnarray}\label{ghDpsi=}
\hat{{\cal D}} \psi^\alpha &:=& d\psi^\alpha - \psi^\beta \wedge
w_\beta{}^\alpha \equiv d\psi^\alpha - \psi^\beta \wedge
(\omega_\beta{}^\alpha + t_1{}_\beta{}^\alpha) \; .
\end{eqnarray}
The Einstein equation of 11D supergravity reads
\begin{eqnarray}\label{cE9a=}
& {\cal E}_{10\, a}:= \frac 1 4 R^{bc}\wedge e_{abc}^{\wedge 8}+ \frac 1 2 \left(i_aF_4\wedge *F_4 +  F_4\wedge i_a *F_4 \right)-  D\psi^\alpha \wedge \psi^\beta \wedge \left(\bar{\Gamma}{}^{(8)}{}_{\alpha\beta}  - \frac 1 2 \bar{\Gamma}{}^{(6)}{}_{\alpha\beta}\wedge e_a\right)   +  \nonumber\\ & +\frac 1 2 \psi^\alpha\wedge \psi^\beta\wedge \left(iF_4\wedge i_a\bar{\Gamma}{}^{(5)}+ i_a\bar{\Gamma}{}^{(2)}\wedge *F_4\right)_{\alpha\beta}  = 0
\; .
\end{eqnarray}
It is important to stress that this equation
does not contain terms of fourth order in fermions \cite{DAuria:1982uck}. In a more standard tensorial form Eq. \eqref{cE9a=} reads
$\left( R_{ac}{}^{bc}-\frac 1 {2}\delta_{a}^{b}R_{cd}{}^{cd}\right)+ \frac 1 {3}\left(F_{a[3]}F^{b[3]}-\frac 1 8\delta_a^b F_{[4]}F^{[4]} \right) \propto \psi\wedge \psi$.

\subsection{Superspace constraints and their consequences}

The advantage of the above first-order action is that it can be lifted to the generalized action of the rheonomic approach (see sec. \ref{GAPsection}),
which implies that all the above equations written in differential forms can be lifted to superspace by the simple prescription in Eq. \eqref{e->E}, complemented by
$A_3(x)\mapsto A_3(Z)=\frac 1 {3!}E^C\wedge E^B\wedge E^AA_{ABC}(Z)$ and $A_6(x)\mapsto A_6(Z)=\frac 1 {6!}E^{C_6}\wedge \ldots \wedge E^{C_1}A_{C_1\ldots C_6}(Z)$.
Analyzing these superform equations we obtain the set of SSP constraints and their  consequences summarized as
\begin{eqnarray}\label{Ta=11D}
T^a&=& -  i E^\alpha \wedge E^\beta \Gamma^a_{\alpha\beta} \; ,
\qquad  \label{Tf=11D}
 T^\alpha=  E^\beta \wedge t_{1\beta}{}^\alpha +
 \frac {1}{2} E^a \wedge E^b T_{ba}{}^\alpha(Z) , \qquad \\
\label{RL=11D}  R^{ab} &=&E^\alpha \wedge E^\beta \left( -{1\over3}
F^{abc_1c_2}\Gamma_{c_1c_2} +  \frac {i}{3^. 5!} (\ast
F)^{abc_1\ldots c_5} \Gamma_{c_1\ldots c_5} \right)_{\alpha\beta}
\; \qquad \nonumber
\\  &&  +  E^c \wedge E^\alpha \left(
-iT^{ab\beta}\Gamma_{c}{}_{\beta\alpha} + 2i T_c{}^{[a \, \beta}
\Gamma^{b]}{}_{\beta\alpha} \right) +  \frac {1}{2} E^d \wedge E^c
R_{cd}{}^{ab}(Z),    \end{eqnarray}
where
\begin{eqnarray}\label{ta:=SSP}
& t_{1\beta}{}^{\alpha}:=E^a T_{a\beta}{}^{\alpha}=  \frac {i}{18} E^a
\left(F_{ac_1c_2c_3}(Z)\Gamma^{c_1c_2c_3} + {\frac 1 8}\;
F^{c_1c_2c_3c_4} (Z)\Gamma_{a c_1c_2c_3c_4}\right){}_\beta^{\;\;\alpha}  \; , \qquad
\end{eqnarray}
is lifting to superspace of $t_{1\beta}{}^{\alpha}$ in Eq. (\ref{t1:=}), and we have the superspace forms
\begin{eqnarray}
\label{cF4=SSP} & {\cal F}_4 := dA_3={1\over 2} E^\alpha \wedge E^\beta \wedge
\bar{\Gamma}^{(2)}_{\alpha\beta} +  \frac {1}{ 4! } E^{c_4}
\wedge \ldots \wedge E^{c_1} F_{c_1\ldots c_4}(Z) \; , \qquad  \\
\label{cF7=SSP}  & {\cal F}_7 := dA_6 + A_3\wedge dA_3= \frac {i}{ 2}
E^\alpha \wedge E^\beta \wedge \bar{\Gamma}^{(5)}_{\alpha\beta} +
 \frac {1}{7! } E^{c_7} \wedge \ldots \wedge E^{c_1} F_{c_1\ldots
c_7}(Z)  \;
\end{eqnarray}
with
\begin{eqnarray}\label{F=*FSSP}
& F_{c_1 \ldots c_7}(Z) = (\ast F_4)_{c_1 \ldots c_7}(Z):=  \frac {1}{4!}
\varepsilon_{c_1 \ldots c_7b_1 \ldots b_4} F^{b_1 \ldots b_4}(Z) \qquad \Leftrightarrow \qquad F_7=*F_4\; .
\end{eqnarray}

The self-consistency of these expressions  (i.e. the torsion and curvature Bianchi identities)  imply that $F_{abcd}=F_{[abcd]}(Z)$ obeys $D_{[a}F_{bcde]}=0$, and that $T_{ab}{}^\alpha (Z)$ and $R_{cd}{}^{ab}(Z)$ are restricted to obey the superfield generalizations of the Rarita-Schwinger and Einstein equations:
\begin{eqnarray}
\label{RS-G8=11DSSP}
& E^b\wedge E^a  \wedge T_{ab}{}^\alpha\Gamma^{(8)}_{\alpha\beta}=0\;, \qquad
\\ \label{EEq=11D=SSP}
& R_{ac}{}^{bc}(Z)-\frac 1 {2}\delta_{a}^{b}R_{cd}{}^{cd}(Z)+ \frac 1 {3}\left(F_{a[3]}F^{b[3]}(Z)-\frac 1 8\delta_a^b F_{[4]}F^{[4]}(Z) \right)=0
\; .
\end{eqnarray}
The superfield generalization of the 3-form  gauge field equations \eqref{d*F4=}, follows from \eqref{F=*FSSP} and the Bianchi identities \eqref{dF7=F4F4}.

\subsection{Killing supervector and  momentum maps in 11D supergravity superspace }

In the case of 11D supergravity the equations of the Killing supervector $ K^A= (K^a, K^{\alpha})$ and corresponding momentum maps will follow not only from the condition of the invariance of the supervielbein 1-form \eqref{DKA=sKilling} and the spin connection superform \eqref{iKR=sKilling}, but also from the requirement of invariance of the 3-form $A_3(Z)$ and its dual $A_6(Z)$, up to their gauge transformations
$\delta A_3= d\alpha_2(Z)$, and $\delta A_6= d\alpha_5(Z)- \alpha_2(Z)\wedge  dA_3$. I.e.
 \begin{eqnarray}\label{vKA3=}
-\delta_K A_3 &=& \imath_K (dA_3) +d\imath_K A_3= d\alpha_{2\; K(Z)} \; , \qquad \\
\label{vKA6=}
-\delta_K A_6 &=& \imath_K (dA_6) +d\imath_K A_6= d\alpha_{5\; K(Z)} - \alpha_{2\, K(Z)}\wedge  dA_3  \end{eqnarray}

Thus introducing, besides the Lorentz momentum map $P_{(K)}{}^{ab}(Z)$  ({\it cf.} \eqref{Pab:=}), the 'electric' and 'magnetic' momentum maps
\begin{eqnarray}  \label{P2K=}  P_{2\; K}:=\alpha_{2\; K(Z)}- \imath_K A_3\; ,  \qquad P_{5\; K}:=\alpha_{5\; K}+A_3\wedge P_{2\; K}- \imath_K A_6\; , \qquad
\end{eqnarray}
we complete the set of equations \eqref{DKA=sK} and \eqref{iKR=sK} by equations for these additional momentum maps following from
\eqref{vKA3=} and \eqref{vKA6=}, respectively,
\begin{eqnarray}  \label{dP2K=}  d P_{2\; K} &=& \imath_K{\cal F}_4 \; , \qquad \;   \\ \label{dP5K=} d P_{5\; K} &=& \imath_K{\cal F}_7 + 2P_{2\; K} \wedge {\cal F}_4 \; . \qquad
\end{eqnarray}

Using the superspace constraints and their consequences collected in Eqs.
\eqref{Ta=11D}--\eqref{F=*FSSP}, we can further specify the above equations arriving at
\begin{eqnarray}\label{DKa=11D}
DK^a&=&2i E^\alpha  \Gamma^a_{\alpha\beta} K^\beta +
 E^{b} P_{(K)b}{}^a \; ,
\\ \label{DKf=11D}
DK^\alpha&=&+ K^\beta t_{1\beta}{}^\alpha + E^\beta (P_{(K)\ \beta}{}^\alpha -\imath_Kt_{1\; \beta}{}^\alpha ) - E^bK^a T_{ab}{}^\alpha\; , \\
\label{dPKab=11D}
DP_{(K)}^{ab}&=& \imath_KR^{ab}= 2E^\alpha \left( - \frac {1}{3}
F^{abc_1c_2}\Gamma_{c_1c_2} +  \frac {i}{3^. 5!} (\ast
F)^{abc_1\ldots c_5} \Gamma_{c_1\ldots c_5} \right)_{\alpha\beta}\,K^\beta
+ E^dK^c R_{cd}{}^{ab}
\nonumber
\\ & &  +E^c \left(
+iT^{ab\beta}\Gamma_{c}{}_{\beta\alpha} - 2i T_c{}^{[a \, \beta}
\Gamma^{b]}{}_{\beta\alpha} \right) \,K^\alpha + E^\alpha \,  K^c \left(
iT^{ab\beta}\Gamma_{c}{}_{\beta\alpha} - 2i T_c{}^{[a \, \beta}
\Gamma^{b]}{}_{\beta\alpha} \right)
,  \\  \label{dP2K=SSP}
dP_{2\; K} &:=&     E^\alpha \wedge
\bar{\Gamma}^{(2)}_{\alpha\beta}K^\beta +  \frac {1}{2 } E^\alpha \wedge E^\beta \wedge E^bK^a
{\Gamma}_{ab\, \alpha\beta}  +\frac {1}{3! } E^{c_4}
 \wedge E^{c_3}\wedge E^{c_2}K^{c_1} F_{c_1\ldots c_4}(Z) \; ,
\end{eqnarray}
and
\begin{eqnarray}
\label{dP5K==SSP}  & dP_{5\; K} =  - i E^\alpha \wedge
\bar{\Gamma}^{(5)}_{\alpha\beta}K^\beta  +  \frac {i}{2^.4! }E^\alpha \wedge E^\beta \wedge E^{c_4}
 \wedge \ldots \wedge E^{c_1}K^{c} {\Gamma}_{cc_1\ldots c_4\, \alpha\beta}  +   \hspace{4cm} \nonumber  \\ & +
 \frac {1}{ 6! } E^{c_7} \wedge \ldots \wedge E^{c_2} K^{c_1}  F_{c_1\ldots
c_7} +  P_{2\; K} \wedge \left( E^\alpha \wedge E^\beta \wedge \bar{\Gamma}^{(2)}_{\alpha\beta} +  \frac {2}{4! } E^{c_4} \wedge \ldots \wedge E^{c_1}  F_{c_1\ldots
c_4}(Z) \right)  \; .
\end{eqnarray}

The natural application of the formalism of Killing supervectors of 11D supergravity is to search for a Komar 9-form of the 11D supergravity
as a closed form in supergravity superspace. This will be one of the subjects of the future publication \cite{IB+PM+TO=D11}.

\section{Conclusion}

In this contribution we have shown, following \cite{Bandos:2023zbs}, how the properties of Killing vectors in supergravity, their supersymmetry transformations
and the supersymmetric generalization of the Killing equations can be deduced from superspace analysis based on the concept of Killing supervector \cite{Buchbinder:1995uq}.
We have used these results to  construct a Noether-Wald charge of $\mathcal{N}=1$ $D=4$ supergravity with fermionic contributions
and Komar charge related to a Killing vector and
its superpartner, the generalized Killing spinor; the latter is diff-, Lorentz- and supersymmetry-invariant (up to a total derivative).
As these charges are apparently  related to the spacetime component approach to supergravity, our method uses essentially
(although this is not explicit in \cite{Bandos:2023zbs}) the so-called rheonomic approach to supergravity \cite{Neeman:1978njh,DAuria:1982uck,Castellani:1991eu},
which provides a solid bridge between the first-order spacetime formulation of supergravity and its superspace formulation.

The results of \cite{Bandos:2023zbs} that we reviewed, provide a solid basis to construct
black hole thermodynamics, although on the level of 'proof of concept' since
black holes with fermionic hair
do not exist in $\mathcal{N}=1$ $D=4$ supergravity \cite{Cordero:1978ud,Gueven:1980be}.
In particular, the Komar charge that we have found seems to suggest the presence of  and additional term proportional to the supercharge in the Smarr formula, but it could simply provide a supersymmetry-invariant definition of mass as well. The no-superhair theorem of \cite{Cordero:1978ud,Gueven:1980be}  prevents us from testing this point in stationary black-hole solutions of this theory. This is why
the practical application of these ideas  requires the generalization of our approach to  ${\cal N}\geq 2$ and/or $D> 4$ supergravity models.
The generalization for minimal ${\cal N}=2$, $D=4$ supergravity will be described in a subsequent publication \cite{IB+PM+TO=N2}. 

In this contribution we have also presented some preliminary
steps of the generalization to the case of 11D supergravity, which can provide a basis for the
thermodynamics of higher dimensional black holes and black p--branes; this  will be the subject of \cite{IB+PM+TO=D11}.

\section*{Acknowledgments}
This work has been supported in part by the MCI, AEI, FEDER (UE) grants PID2021-125700NB-C21 (“Gravity, Supergravity and Superstrings” (GRASS)) (IB and TO),
PID2021-123021NB-I00 (PM) and IFT Centro de Excelencia Severo Ochoa CEX2020-001007-S (TO), by FICYT through the Asturian grant SV-PA-21-AYUD/2021/52177 (PM)
and by the Basque Government Grant IT1628-22 (IB). TO wishes to thank M.M. Fern\'andez for her permanent support.

\end{document}